\documentclass[  
twoside, 
openany, 
reqno]{amsart}        
\usepackage[english]{babel}         
\usepackage{graphicx}
\usepackage{hyperref} 
\usepackage{verbatim}  
\usepackage{float}   
\usepackage{pdfpages}

\begin{document}   
	\setlength{\parindent}{0pt}   
	\def\sangria{\hskip 15pt\relax}        
	\renewcommand{\bibname}{Referencias}                    
	
	\title[]{A REPONSE TO THE SHELDON GOLDSTEIN OBJECTION TO A SERIES OF COVARIANT RELATIVISTIC BOHMIAN MODELS.}
	
	\begin{center}
	{\large Sergio Hern\'andez-Zapata and Gerardo Ruiz-Chavarr\'ia} \\ 
	{\small Departamento de F\'isica; Facultad de Ciencias; UNAM, CDMX, M\'exico} \\ 
	{\small sergiohz@ciencias.unam.mx (2026)} 
	\end{center}
	
	\maketitle

	\begin{abstract} 	In this work we try to answer a very important objection to the Covariant Bohmian Models type Nikoli\'c and Hern\'andez-Zapata that was formulated by the american physicist Sheldon Goldstein (private communication, 2010) about the imposibility to interpret the configuration space-time density found by the authors of these models like a probability density. The problem is that it is very dificult to show that the $N$ space-time configuration density has a finite integral. That is very important in order to normalize the density to produce a PDF (probability density function). All the steps in the demonstrations seemed complete but this objection must be resolved imperatively before to continue using this kind of models with full confidence. Since the aim of Bohmian Mechanics (also known as the De Broglie-Bohm Mechanics) is to provide a precise and objective formulation of Quantum Mechanics free from vagueness and obscurity (Bell point of view at least), addressing this issue is crucial. In this article, we set out to tackle this by employing a specific type of function that we call an "Arrangement Function".
		
	
	\medskip
	
	\noindent\textbf{keywords:} Bohmian Mechanics, Space-Time Density, Nikoli\'c Time, Arrangemente Function.
	\end{abstract}

	\section{Introduction}
	
	\sangria Bohmian Mechanics (also known like the De Broglie-Bohmian Theory) is a Completion of Quantum Mechanics. It introduces some new elements that are not present in the original formulation. At the beginning it was born like an Explanation of a Very Powerful Formalism that was absolutely invincible in making correct experimental predictions. That was the Monumental Work of Niels Bohr and a very large number of leading scientists from Gotinga, Copenhague and a lot of places around the World. The search of Albert Einstein, Louis De Broglie, David Bohm, John Bell between others was to find a manner to see this Powerful Tool in a way independent of Observers. That is to say, the Search was to integrate Quantum Mechanics to the Espirit of all the Science and the Illustration. To describe the Nature in an Objective Manner. The Idea is that it is irreductible to the Human Representation and Human Practice. In Mexico, by example, there is the Scholl of Ana Mar\'ia Cetto and Luis de la Pe\~na who tried to form innumerable generations in this Espirit [\cite{37}, \cite{38}]. John Bell, the Irish scientist, known the seminal papers of David Bohm and he saw that it was a good counterexample of almost all Non-Hidden Varible Theorems including John von Neumann's. He studied with a great profundity the principal characteristics of Bohmian Mechanics and discovered that this Theory has a very peculiar thing. It was Non-Local, that is to say, two particles with a spacelike separation could interact each other. That was a very difficult thing to accept in that time, because Theory of Relativity seemed to point in other directions. To study all this problematic in a very profound way and to avoid extend this paper unnecessarily  we recomend the Jean Bricmont's papers and books about Quantum Mechanics and Bohmian Mechanics [\cite{13}, \cite{14},\cite{15}, \cite{16}, \cite{17}, \cite{18}, \cite{19}, \cite{20}, \cite{21}, \cite{22}, \cite{23}, \cite{26}, \cite{27}, \cite{28}, \cite{31}, \cite{32}] and all the papers of the Rutgers School (In all this paper we call this School 'Bell Scholl' like it be point later) [\cite{4}, \cite{6}, \cite{8}, \cite{9}, \cite{10}, \cite{11},\cite{36}] and of course all that the proper John Bell has written about Quantum Mechanics, by example [\cite{7}].  
	
    \sangria Around 2009-2010 some papers were published that exposed some models of Relativiic Bohmian Mechanics that had the property of being Lorentz Covariant besides to be Non-Local. The first published paper was Hrvoje Nikoli\'c one based on Klein-Gordon equation. Sergio and Ernesto Hern\'andez-Zapata worked a Klein-Gordon model whose central philosophy and central ideas, with some differences, would have been published by Nikoli\'c. The objective of the Hern\'andez-Zapata and Hern\'andez-Zapata work was to study if this Klein-Gordon Model produced good and reasonable Clasical ($ \hbar \rightarrow 0 $) and Non-Relativistic Limits  ($ c \rightarrow \infty $ ) In the first objective was very important to take like a model the procedure of Valia Allori et al. to take the Classical Limit, they worked Non-Relativistic Bohmian Mechanics and Hern\'andez-Zapata et al. extended these ideas to the Relativistic  Covariant Klein-Gordon Model. They suppose that the particles were subjected to an external Electromagnetic Field given by $ A_\mu $ in order that the system be more interesting. (That is to say the system continues to be some kind of Ideal Gas because the particles do not interact electromagnetically between each other). They obtain results very similar to Allori et al. but in the Relativistic Case. In this case, the parameter $ \sigma $ that parameterize the particles trajectories tends to be the proper time in the limit when $ \hbar \rightarrow 0 $ and each particle is subjected to a Lorentz Force. The Dynamics then is completely Local and coincides with a Classical Einsteinian Dynamics, that is to say, at the moment of taking Classical Limit all the Non-locality disapears, maybe it rests in the initial condition but even there is now hidden. Locality emerges when they take the Classical Limit, exactly as it was the case in the Non-Relativistic Limit studied by Allori et al.

    
    
    \sangria By other side, they proceed to study the Non-Relativistic Limit. This last Limit has been studied a lot at least for the Wave Function. In this case they tried to cover also the Bohmian aspect of the Non-Relativistic Limit. It was found following a procedure based in the habitual studies of Bell School (Goldstein-D\"urr-Zangh\'i) that the Configurational Times could be resolved independently of the other Configurational coordinates but there was not really a manner in which they became exactly the same Universal Time (Like, it is typical of Non-Relativistic Physics) because they appear constants that can not be really supprimed. This was an interesting point because it opened some ways to explore other possibilities. It was the most consistent paper thay they publish that year.

    
    \sangria As is well known to everyone, the Dirac Equation is an equation that although distict from the Klein-Gordon equation is extremely important for the study of the electrons. It could be interesting to attack it with the previous ideas. Sergio Hern\'andez-Zapata works this problem [7]. One aspect crucial was to obtain a Scalar that could be a good candidate for Space-Time Density. The problem was that the more direct Scalar that was produced by the Dirac Equation was not positive definite. Then he took the absolute value of this Scalar like his propose. The only mechanism that he could take at that moment to explore if the propose was good consisted of following the methods that he has learned in [\cite{4}] to obtain the Classical Limit. He only obtained a clean Classical Limit if he formulated a hypothesis about the phases of the spinor.
    
     \sangria If he supposes that $ \hbar \rightarrow 0 $ in the Dirac Equation and, he works the Classical Limit like Allori et al., then, under certain hyphotesis, a Classical Limit governed  by the Relativistic Hamilton-Jacobi Equation and the Equation that relates the Velocity four vector with the four vector gradiente of the Classical Action by means of multiplying by the inverse of the mass is obtained. The problem is that he only obtains it if he did the Hypothesis that in the Classical Limit (i.e., when $ \hbar \rightarrow 0 $) the phases of the spinor becomes equal to each other.

     \sangria Sergio Hern\'andez-Zapata try to justify it, because how he obtains a very Clean Limit, he thought that this phenomenon had to be real, the Classical Action had to be that common limit of the Spinor Phases. Something like an strange border between the Microscopic World and the Classical World. An interesting aspect, that was not addressed in the paper is that if he does with the Spinor of a Many particles System and he take the Classical Limit in the sense of Allori et al. he obtained  a  very similar system to the obtained in the Klein-Gordon system. It can be utilized the same m\'ethod that in the Klein-Gordon Equation and we obtain directly the Einsteinian Classical trajectories of particles esencially disentangled subjected to an external Eletromagnetic Field given by the four vector potential $ A_\mu $. In that case, it can be obtained again the typical trajectories of a Classical Einsteinian Mechanics. Everything seemed to work very well. There is something that he did not analize and maybe if he would have done so he could not produce similar results. It is a very strange thing that any restriction  in the interactions in the space-time is not present in the original Models. That is to say, any place in the space-time can affect any other place. But at least the Classical Limit and the Non-Relativistic Limit were very Clean. The paper about Klein-Gordon was published in Foundations of Physics and everything seemed to work okay. Then, he send the article to the american physicist Sheldon Goldstein. He have had illustrated him about some points related to the paper of Allori [\cite{4}]. After, he send to him the paper his response left him devastated. In his email, Sheldon Goldstein points out a dificulty that seemed insurmountable. The Space-Time Density that was proposed in our papers (Nikoli\'c, Hern\'andez-Zapata and Hern\'andez-Zapata) was not integrable. This meant, that he had  suposed like realized steps, like taking the Conditional Wave Function, for example, in the Classical Limit, when he substituted the Configurational Times in the Wave Equation, nothing was realy valid. To reply the objection of Sheldon Goldstein was crucial to validate all the Models of this type. This problem had to be  adressed. This is the topic of this paper. In order to adress this topic , we need a new  concept called an 'Arrangement Function' that I will describe in the development of the paper. 
     		
     \sangria The paper is organized as follows. In the first section, first subsection, we describe the Bohmian Model based on the Klein-Gordon Equation and study the Non-Relativistic Limit. In order to understand the discusion is very important to take into account all the time the difference between generic and configurational variables. In the second subsection, we introduce our first Arrangement Function. This is very important because basically restricts all interactions to be Spacelike interactions. Then, we make abstraction of the Sheldon Goldstein convergence objection and we show that we obtain in the Non-Relativistic Limit the usual Non-Relativistic Bohmian Mechanics with a universal Configurational Time $ T $.

     \sangria In the second subsection, we describe the Bohmian Model based on the Dirac Equation. We mantain the discusion restringed to two entangled particles. The generalization to many particles is straighforward. We take also in this case the Non-Relativistic Limit. We use the same Arrangement Function that in the case of Klein-Gordon Equation and then esentially we obtain the same results that in the Klein-Gordon case. 
     We study two Non-Relativistic Limits, when $ c \rightarrow 0 $ and when the spatial Configuration separation of any two particles divided by $ c $ is negligible with respect to the precisi\'on of our Clocks. We use another class of Arrangement Function whose 'Support', the domain in which the Arrangement Function is not negligible and has value 1. The 'Support' is esentially the gap between the level curves 0.0 and 1.2 of The Interval Function. It is essentially a preparation to give a reponse to the Sheldon Goldstein Objection.

     \sangria At the end, we use an Hypothesis about the existence of Timelike trajectories (Geodesics in space-time) like the domains in which a Dinamics happens. A massless particle moves over each Timelike trajectory and its Probability Density Function is given by $ F( \tau ) $ then $ F( \tau ) d \tau $ is the probability that at Nikoli\'c time $ \sigma $ the particle is in the proper time interval $ d \tau $.
     
     \sangria At last, we use all the concepts constructed in this paper to define an entanglement Dynamics between a particle moving in Space-time $ x_{\mu} $ and a particle moving in a TimeLike curve with parameter $ \tau $ and we study the Space-Time-$ \tau $ Density and we argument that the integral of this Density is finite.

    \section{SPACELIKE INTERACTION HYPOTHESIS}  
	
	\subsection{The Case of a Klein-Gordon Wave Equation}
	
	\sangria  Bohmian Mechanics is an extension of Quantum Mechanics in which the particles have real positions at all times and follow trajectories. So, in addition to the Wave-Equation, a set of equations is added that describe how the particles are propelled by the Wave Function. In the Non-Relativisticcase , it has been proven that the predictions of this Theory are identical to those predictions of Ordinary Quantum Mechanics. In fact, Bohmian Mechanics is a good counterexample to many hidden variable theorems. This is why the search for Relativistic versions of it is very important. In this article we study an extension of Bohmian Mechanics to Special Relativity [\cite{1},\cite{2}, \cite{3}], which does not use a foliation (references to study the idea of foliation to search in Relativistic Bohmian Mechanics [\cite{24}, \cite{25}, \cite{29}, \cite{30}, \cite{33}, \cite{34}, \cite{35}]) but a spatio-temporal density so that the particles are in quantum equilibrium throughout space-time for every value of a parameter $\sigma$, introduced by Hrvoje Nikoli\'c. We are making abstraction in this description of Sheldon Goldstein objection. This parameter is reduced to the proper time of particles in the Classical Limit, that is, when $\hbar$ is much smaller than all characteristic actions of the system [\cite{2}, \cite{3}]. (In order to more broadly understand what Bohmian mechanics is about we recommend papers [\cite{4},\cite{6},\cite{7}]).

	\sangria There are certain kind of Bohmian Models that are Lorentz-Covariant, that is, the Wave Equation and the Guiding Equations are transformed covariantly in the different frames of reference of Special Relativity. In this paper, we study one model based on the Klein-Gordon equation [\cite{1},\cite{2}], and another based on the Dirac Equation [\cite{3}]. Both Models can be extended to several particles. In that way we have Bohmian Mechanical Models that are covariant and at the same time they are Non-local. This kind of Model was introduced by Hrvoje Nikoli\'c for the Klein-Gordon Equation \cite{1} and have some peculiar properties: In the first time the Wave Function depends on the generic coordinates $ {\bf x}, ict  $ (The generic variables are denoted with small letters) of all particles. Like the generic times of all particles are independent the formalism is multitemporal. As is usual in Bohmian Mechanics, the dynamics of the configurational variables $ {\bf X}, icT $ (The configurational variables are denoted with capital letters) is determined by a Guide Equation that describes how the Wave Function guides the dynamics of the System.  There is then a $4N$-dimensional velocity field  where the four-velocity of each particle transforms like a four-vector. The path of the system of all particles follows the field parameterized by a scalar time $\sigma$, introduced in [\cite{1}] for this type of models. The streamlines of the field are then the set of universal lines of the particles. Another characteristic of the Models is that a $N$-spatial-time density of probability (We make abstraction of the Sheldon Goldstein objection by now) is introduced and then a Global Continuity Equation is satisfied. Therefore, if the Configuration System is in quantum equilibrium in all $N$-space-time at $\sigma = \sigma_0$ it will be in quantum equilibrium for all $\sigma$. 
	
	\sangria A new idea to attack the problem
	
	\sangria In the reference \cite{5} the following system of Equations is studied:
	
	\begin{equation}\label{tomonaga1}
	\{ H_n ({q_n},{p_n},\mho ({q_n}_,{t_n})) + \frac{\hbar}{i} \frac{\partial}{\partial t_n} \} \Phi \left(
	{q_1}{t_1},{q_2}{t_2},...,{q_N}{t_N} \right) = 0   
	\end{equation}
	
	where $n=1,2,...,N$. By other side, $\mho ({q_n}_,{t_n}))$ comes from the external forces to which the $n$th particle is subjected. A condition in order to solve the system of equations \eqref{tomonaga1} is that the following $N^2$ equations  be satisfied
	
	\begin{equation}
	\left( H_n H_{n'}-H_{n'} H_n \right) \Phi \left( {q_1}{t_1},{q_2}{t_2},...,{q_N}{t_N} \right)  = 0.
	\end{equation}

	
	
	\sangria By the other side, if the conmutators $H_n H_{n'}-H_{n'} H_n $ are equal to $0$ for every $n, n'$ the solution of the system of Equations \eqref{tomonaga1} exists. This happens if
	
	
	$$ |q_n - q_{n'}|^2 - c^2 \left( t_n - t_{n'} \right) ^2 \geq 0. $$
	
	
	\sangria In the following we present a redefinition of the Born-Nikoli\'c rule and the Bohmian Guide Equation by means of the Introduction of certain Arrangement Functions. A description of the procedure that I utilized in the following is to write the Continuity Equations and multiply the argument for an Arrangement Function and its Inverse, that is to say, by one. The Arrangement Function is a Function that in some 'Support' it equals 1 and outside that domain its value is negligible compared to 1, yet it is always continuous and infinitely differentiable and positive. We will provide examples to show that this is possible. The Arrangement Function is incorporated to the $N$-space-time density and The Inverse of the Arrangement Function is incorporated to the Bohmian Guide Equation. We have then a System very similar to the Original for the Klein-Gordon Equation. What is changed? The Arrangement Function is exactly one if it is an Arrangement Function properly constructed, in a certain Region. Outside this Region it is negligible but positive. In principle it must be a rigid procedure, in the sens of Weinberg [\cite{12}], in order to construct this type of Arrangement Functions. We have not produce something like that. Only, we are looking for the objective that all the description be Covariant. With this restriction what we have to do is to propose with some simple criteria an Arrangement Function. That it behaves pretty well in a certain Region and Outside this Region but, maybe, with some problems in the Transit Region between some domains. All that will be realized in order to obtain a Bohmian type Model, for the Klein-Gordon Equation or for the Dirac Equation that let be Lorentz Covariant and in the Relevant Region that let be one, Outside this Region that let be negligible but positive and with all derivatives continuos $ C^{(\infty)} $.
	

	\section{NON-RELATIVISTIC LIMIT TAKEN UNDER \textit{“THE SPACELIKE INTERACTION HYPOTHESIS”}}

	\sangria Inspired by the Tomonaga arguments we think that the Tomonaga Model sugests something interesting. While the 2010 Bohmian Relativistic Models (Nikoli\'c, Hern\'andez-Zapata and Hern\'andez-Zapata) do not have any restrictions, in the Tomonaga Model there are a very important restriction. At the end, all interactions are spacelike. That is to say, the Quantum Interactions seem to act in directions that are prohibited in  Classical Physics. In Classical Physics all interactions are timelike interactions, then we want to explore the possibility that, the Quantum Interactions, entanglement and interactions like that be strictly Spacelike. That is the idea that Tomonaga inspired to us. That is the first step to construct an Arrangement Function. This will be our first example of the procedure to arrange the Relativistic Bohmian Model using this kind of procedure based on Arrangement Functions. Another thing, we think that we will not use the Hamiltonians used by Tomonaga. We use only Fields. In this article, we think of all the particles subjected to the same Electromagnetic External Field determined by a four-vector potential $ A_\mu $.
	We have practically Models like that in the 2010 papers. We must modify these in order to obtain the restriction that we are interested in. The first step is to introduce an Arrangement Function in this line of Thought. The idea is then to solve the Wave Equations in all $N$-space-like configuration and then to Arrange the things to obtain only spacelike interactions. We proceded to the Construction of The Corresponding Arrangement Function to Obtain Only Spacelike Interactions.

	\subsection{CONSTRUCTION OF THE CORRESPONDING ARRANGEMENT FUNCTION TO THE SPACELIKE INTERACTIONS}

	\sangria We will consider the following Function:
 	
	\begin{equation} \label{exp}
	F(u)=\left\{
	\begin{array}{lr}
	exp\{\frac{-a}{u^2}\} & \mbox{if } u > 0 \\
	0 & \mbox{otherwise}
	\end{array}
	\right.
	\end{equation}

	\sangria This Function is $C^{(\infty)}$ class ever it is not Analitical. Its value is $0$ for $u \leq 0$ and it tends to $1$ when $u \rightarrow \infty $.

	\sangria Now we construct the Following Function that has some interesting characteristics. 

	$$ G_{(i, j)} = 1 - F(-u) $$

	$$ \mbox{where} \, \, \, \,  u =  ||x^{(i)}-x^{(j)}||^2-c^2(t^{(i)}-t^{(j)})^2  > 0 $$ 

	\sangria This Function is $ C^{(\infty)} $, positive and it has the properties that we are looking for.
	
	
	\sangria This new Function has the Value $1$ when the interval between the two events $ {x_{\mu}}^{(i)}$ y $ {x_{\mu}}^{(j)} $ is spacelike. It is positive but negligible when the Interval is Timelike. And there is effectively a transition between the two situations mentioned before in the light cone. The important thing is that the Function $ G_{(i, j)} $ is positive. This is very important because we must divide by this Function in the subsequent experiments. The Function $ G_{(i, j)} $ is our first example of an Arrangement Function.

	\sangria The graph of the Function \eqref{exp} $ F(u) $ is shown in the following figure where $ a = 30 $ in order only to exemplify. It can be proven that this Function, while not analytical, is continuous and has derivatives of all orders in all its domain and these derivatives are also continuous.

	\hspace{3 cm}

	\includegraphics[scale=0.5]{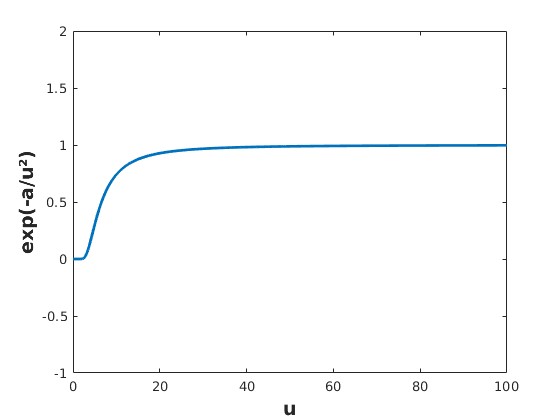}

	\sangria When $u/a$ is great enough function \eqref{exp} is indistinguishable from $1$. When  $0< u/a << 1$ the function is indistiguishable from $0$ and there it has all its derivatives and these are continuous. The function is of class $C^{(\infty)}$ but it is not analytical. Consider the following function now:

	$$H(x_\mu^{(1)}, x_\mu^{(2)}, \dots , x_\mu^{(n)})=\prod_{i \neq j} G_{(i,j)} (x_\mu^{(i)}, x_\mu{(j)}).$$

	Then we look for solutions $\Psi$ of the wave equation in the configurational spacetime of the form

	\begin{equation}\label{10}
	\Psi(x_\mu^{(1)}, x_\mu^{(2)}, \dots , x_\mu^{(n)})  =H(x_\mu^{(1)}, x_\mu^{(2)}, \dots , x_\mu^{(n)})\varphi(x_\mu^{(1)}, x_\mu^{(2)}, \dots , x_\mu^{(n)})
	\end{equation}	

	where the function $ \varphi $ is a solution of the partial Wave Equations in a Bohmian system like Nikoli\'c, Hern\'andez-Zapata and Hern\'andez-Zapata (Klein-Gordon and Dirac Equations are ones of the published Models). 

	\subsection{THE CASE OF A KLEIN-GORDON WAVE EQUATION}

	\sangria In the case based on the Klein-Gordon Equation, we consider the Scalar Wave Function   $\Psi({\bf x}^{(1)}, c t^{(1)}, {\bf x}^{(2)}, c t^{(2)},\dots, {\bf x}^{(N)}, c t^{(N)}) $ for $N$ spinless particles. By other side, we suppose that the particles interact with an external Electromagnetic Field $A_\mu$ but not between each other electromagneticaly. We are using a multitemporal formalism since time and space coordinates are treated in the same manner, that is, each particle has four associated coordinates; three space coordinates and one time coordinate. In total, we have $N$ time coordinates, one for each particle [\cite{1},\cite{2},\cite{3},\cite{5}]. The Wave Function satisfies $N$ Klein-Gordon equations, one for each particle of the form
    
    \begin{equation}\label{kleingordon}
    \left( \frac{\hbar}{i} \frac{\partial}{\partial {{x_\mu}^{(i)}}} -\frac{e}{c}{A_\mu}^{(i)}  \right) \left(\frac{\hbar}{i} \frac{\partial}{\partial {{x_\mu}^{(i)}}} -\frac{e}{c}{A_\mu}^{(i)}  \right) \Psi  = -{m_i}^2 c^2 \Psi
    \end{equation}
    
    $i=1,..,N$ where $\Psi$ is a Complex Scalar Function. If we write the Wave Function in polar form $\Psi = R e^{iS/\hbar}$ and then separate the real and imaginary parts, we obtain that the Klein-Gordon Equation is equivalent to the two equations \eqref{kleingordonreal} and \eqref{kleingordonimaginario}.
    
    \begin{equation}\label{kleingordonreal}
    \left( \frac{\partial S}{\partial {{x_\mu}^{(i)}}} -\frac{e}{c}{A_\mu}^{(i)}  \right) \left( \frac{\partial S}{\partial {{x_\mu}^{(i)}}} -\frac{e}{c}{A_\mu}^{(i)}  \right)  = -{m_i}^2 c^2 + \frac{\hbar^2 {\square^2}_{(i)}R}{R}
    \end{equation}
    
    \begin{equation}\label{kleingordonimaginario}
    \frac{\partial}{\partial {{x_\mu}^{(i)}}}  \left( R^2  \left( \frac{\partial S}{\partial {{x_\mu}^{(i)}}} -\frac{e}{c}{A_\mu}^{(i)}  \right) \right)=0.
    \end{equation}
    
    \sangria The equation \eqref{kleingordonreal} is very similar to the Relativistic Hamilton-Jacobi equation except for the term $ {\hbar^2 {\square^2}_{(i)}R}/{R} $, $S$ stands for the action. The equation \eqref{kleingordonimaginario} is a continuity equation for a steady flow in a four-dimensional space.   It can then be proven that the global continuity equation is satisfied
    
    \begin{equation}\label{globalcontinuidad}
    \frac{\partial {R^2}}{\partial\sigma} +  \sum_{i=1}^{N} \frac{\partial}{\partial {{x_\mu}^{(i)}}}  \left( \frac{R^2}{m_i}  \left( \frac{\partial S}{\partial {{x_\mu}^{(i)}}} -\frac{e}{c}{A_\mu}^{(i)}  \right) \right)=0
    \end{equation}
    
    \hspace{1cm}
    
    since $R^2$ does not depend on $\sigma$ and all equations \eqref{kleingordonimaginario} have been divided by the mass of the corresponding particle and added. Setting $ {V_\mu}^{(i)} = \frac{1}{m_i} \left( \frac{\partial S}{\partial {{x_\mu}^{(i)}}} -\frac{e}{c}{A_\mu}^{(i)} \right) $ the equation \eqref{globalcontinuidad} can be written as
    
    \begin{equation}
    \frac{\partial {R^2}}{\partial\sigma} +  \sum_{i=1}^{N} \frac{\partial}{\partial {{x_\mu}^{(i)}}}  \left( R^2 {V_\mu}^{(i)} \right)=0.
    \end{equation} 
    
    \hspace{1cm}
    
    \sangria In Bohmian Mechanics the Wave Equation is complemented by another equation called the Guiding Equation that tells us how the dynamics of particles is guided by the Wave Function. This last equation has the following form in the Model.
    
    \begin{equation}\label{EcuacionGuia}
    \frac{{dX_\mu}^{(i)}}{d \sigma}  = \frac{1}{m_i} \left( \frac{\partial S}{\partial {{X_\mu}^{(i)}}} -\frac{e}{c}{A_\mu}^{(i)}  \right). 
    \end{equation}
    
    where $\sigma$ is a scalar time that is used to parameterize the trajectories of particles in the $N$-space-time \cite{1}, we call this parameter Nikoli\'c time. On the other hand, the right side of the equation \eqref{EcuacionGuia} is evaluated in the spatio-temporal configurational coordinates of all the particles.
    
    \sangria It follows that the idea of Quantum Equilibrium can be extended to this Model. If the system of particles is distributed in the $N$-space-time with probability density proportional to $R^2$ at time $\sigma = \sigma_0$ it will continue to be so distributed for all time $\sigma$. The classical and Non-Relativistic limits of this model were studied in [\cite{2}]. The argument for why this happens is exactly the same as the typical argument for Quantum Equilibrium in typical Non-Relativistic Bohmian Mechanics. Since $R^2$ is a scalar and the volume element in the $N$ spacetime is invariant in Special Relativity, since the Jacobian of the Lorentz transformations is 1 in the entire $N$-spacetime \eqref{volumeninv}, it follows that the Quantum Equilibrium condition is valid in all inertial reference frames for every value of $\sigma$.
    
    \begin{equation}\label{volumeninv}
    \frac{\partial \left(x',y',z',ct'\right)}{\partial \left(x,y,z,ct\right)} = 1
    \end{equation}
   	
   	\sangria In order to obtain the Non-Relativistic limit we introduce a new phase  $\tilde{S}$ defined by
	
    \begin{equation}
	\tilde{S} \equiv S + {c^2} \sum_{j=1}^{N}  {m_j} t^{(j)}. 
	\end{equation} 
	
	\sangria	We had taken [\cite{2}] the Non-Relativistic limit, that is to say when all the characteristic velocities are much lesser than the light velocity, of the $N$ Klein-Gordon Equations and we got $N$ Non-Relativistic Equations, Schr\"odinger type, with respect to the generic coordinates of each particle. 
	
	\begin{equation}\label{Schrodi}
	i\hbar\frac{\partial \tilde{\Psi}}{\partial t^{(i)}}=-\frac{\hbar^2}{2m}\nabla^{(i)2}\tilde{\Psi}+e^{(i)}\Phi^{(i)}\tilde{\Psi}
	\end{equation}
	
	where $\tilde{\Psi} = R e^{\frac{i \tilde{S}}{\hbar}}$. 
	
	\sangria One solution of the system of $N$ equations \eqref{Schrodi}, one for each particle, yields the Wave Function for this Non-Relativistic System of Equations.
		
	\sangria In the Non-Relativistic Limit we had in [\cite{2}] that the Configurational time coordinates and the configurational space coordinates decouple.  We obtain that the derivative of each configurational time coordinate with respect to $\sigma$ parameter is 1 in this limit. Therefore, this implies the following identities
	
    \begin{equation}\label{tiempos}
	T^{(i)} = \sigma  +  \delta^{(i)}   ;  i = 1, 2, \dots , N.
    \end{equation}

    \sangria Now we substitute the Configurational times in the Wave Function. According to the typical arguments of Bohmian Mechanics, when a set of configurational variables in the Guiding Equations is decoupled, and you solve the equations for these variables, then when you substitute the respective Configurational Variables in the Wave Function you obtain the Conditional Wave Function for the variable that remains generic. [\cite{8},\cite{9},\cite{10},\cite{11}]. Besides, the obtained subsystem is in Quantum Equilibrium. We still have a Multitemporal System, but now with Configurational time variables. What we do is substituting the configurational times \eqref{tiempos} in this solution and we obtain a Wave Function that depends on the space coordinates of the particles and depends on $\sigma$ through the time $T^{(i)}$. That is to say, we obtain
   
	$$\tilde{\Psi} (x^{(1)}, x^{(2)}, x^{(3)}, \dots , x^{(n)}, \sigma  +  \delta^{(1)},  \sigma  +  \delta^{(2)}, \dots , \sigma  +  \delta^{(n)})$$

	where we have changed the order of the independent variables.
	
	\sangria That is exactly that we did in the paper [\cite{2}]. The problem is that this analysis is wrong because in order to be correct it is necesary that a PDF (Probability Density Function) can be defined. That is to say, it is necesary that the $N$-space-time configuration density be integrable. That is to say, the integral of this density must be finite (Sheldon Goldstein objection to the 2010 Relativistic Bohmian Models, Nikoli\'c, Hern\'andez-Zapata and Hern\'andez-Zapata). We must change some things in order to improve the Models. 
	
	 \sangria We remember that the new hyphotesis is that esentially only if the separation of two particles at Nikoli\'c time $ \sigma $ is Spacelike there is interaction between the particles. 
			
	
	
	\sangria For the Klein-Gordon Equations for $ N $ particles the Continuity Equations are
	
	\begin{equation}
	\frac{\partial}{\partial {{x_\mu}^{(i)}}}  \left( R^2  \left( \frac{\partial S}{\partial {{x_\mu}^{(i)}}} -\frac{e}{c}{A_\mu}^{(i)}  \right) \right)=0.
	\end{equation}.
	
	\sangria By the other side, in order to impose restrictions we use an Arrangement Function that impose that the Typical (In the same sense that Statistical Mechanics [\cite{39}, \cite{40}]) Configuration (The extremely most probable Configuration of the particles is space-like). Then we use the Following Arrangement Function
	
	  \begin{equation}
	  G_{(i, j)} = 1 - F(-u) \label{rococo}
	  \end{equation}
	
	$$ \mbox{where} \, \, \, \, u =  ||x^{(i)}-x^{(j)}||^2-c^2(t^{(i)}-t^{(j)})^2  > 0 $$. 
	
	\sangria Now, we arrange the things (It is the reason for the name 'Arrangement'). 
	
	\sangria When $u/a$ is great enough function \ref{rococo} is indistinguishable from $1$. When  $0<u/a << 1$ the function is indistiguishable from $0$ and there it has all its derivatives and these are continuous. The function is of class $C^{(\infty)}$ but it is not analytical. Consider the following function now:

	$$H(x_\mu^{(1)}, x_\mu^{(2)}, \dots , x_\mu^{(n)})=\prod_{i \neq j} G_{(i,j)} (x_\mu^{(i)}, x_\mu{(j)}).$$
	
	\sangria We arrange the Continuity Equations in the following manner
	
	\begin{equation}
	\frac{\partial}{\partial {{x_\mu}^{(i)}}}  \left( H R^2 H^{-1} \frac{1}{m_i}  \left( \frac{\partial S}{\partial {{x_\mu}^{(i)}}} -\frac{e}{c}{A_\mu}^{(i)}  \right) \right)=0.
	\end{equation}.
	
	The arrange $ N $-space-time density is  
	
	$$ \rho_{\text{$N$-space-time}} = H R^2 $$
	
	and the arrange Bohmian Guiding Equations are
	
	$$ \frac{ d X^{(i)}_\mu }{ d \sigma } = H^{-1} \frac{1}{m_i}  \left( \frac{\partial S}{\partial {{x_\mu}^{(i)}}} -\frac{e}{c}{A_\mu}^{(i)}  \right) $$
	
	\sangria If we work inside the 'Support of the Arrangement Function $ H $' then $ H^{-1} = 1 $ and the analysis is absolutly equal to the analysis in the paper \cite{2} . Then, we obtain the same Schrödinger Equations that they  obtain in the paper:
	
	\begin{equation}\label{2}
	i\hbar\frac{\partial \tilde{\Psi}}{\partial t^{(i)}}=-\frac{\hbar^2}{2m}\nabla^{(i)2} \tilde{\Psi}+e^{(i)}\Phi^{(i)} \tilde{\Psi}
	\end{equation}
	
	and the Guiding Bohmian Equations 
	
	$$ \frac{ d \vec{X_i} }{ d \sigma } = \frac{1}{m_i} {\nabla_{i}}{ \tilde{S} } $$
	
	for the Configurational Times the Bohmian Guiding Equation are
	
	$$ \frac{d T^{(i)}}{d \sigma}  = 1 $$
	
	All the details are in the paper \cite{2}. The clear thing is that the configurational times are decoupled from the configurational space positions of the particles. A very important observation, the new $N$-space-time density continues to be Non-Integrable but like the new 'Support of the Density is reduced' we are in the correct way. We are going to make abstraction of The Sheldon Goldstein Objection. In the first place, we integrate to obtain the Configurational times $ T^{(1)} $, $ T^{(2)} $, ..., $ T^{(N)} $ obtaining the following relations
	
	$$ T^{(i)} = \sigma + \delta_i $$
	
	$ i = 1, 2, ..., N $. But we must remember that the Configuration is a typical Configuration (In the sense of Statistical Mechanics), and then in the way of taking the Non-Relativistic Limit the interval between the Configurational Coordinates of any two particles is spacelike too. And then, for two any two particles
	
	$$ \frac{\lvert \vec{X}_{(i)} - \vec{X}_{(j)} \rvert}{c} \ge  \lvert T^{(i)} - T^{(j)} \rvert $$
	
	Like we are taking the Non-Relativistic Limit, that is to say $ c \rightarrow \infty $, in that Limit the left hand side of the equations $ \rightarrow 0 $. And then in that limit  $ T^{(i)} = T^{(j)} $ for any $i, j$. That is to say
	
	$$ T^{(1)} = T^{(2)} = ... = T^{(N)} = T $$
	
	All the Configurational Times colapse on the same value in the Non-Relativistic Limit.
	
	\sangria Considering the sum of the left side of each of these equations, ignoring the constant $i\hbar$, \eqref{2}, we have that:
	$$\sum_{i=1}^{N}\frac{\partial \tilde{\Psi}}{\partial T^{(i)}}=\sum_{i=1}^{N}\frac{\partial \tilde{\Psi}}{\partial T^{(i)}}\frac{\partial T^{(i)}}{\partial T}$$
	
	since $\frac{\partial T^{(i)}}{\partial T}=1$ and because of the chain rule we obtain
	
	\begin{equation}\label{3}
			\frac{\partial \tilde{\Psi}}{\partial T}=\sum_{i=1}^{N}\frac{\partial \tilde{\Psi}}{\partial T^{(i)}}.
	\end{equation}

	And analyzing the sum of the right side of the equations:

	\begin{equation}\label{4}
	-\hbar^2\sum_{i=1}^{N}\frac{\nabla^{(i)2}\tilde{\Psi}}{2m_i}
	\end{equation}

    is obtained by adding the free terms of the Hamiltonian.

	\begin{equation}\label{5}
	\sum_{i=1}^{N}e^{(i)}\Phi^{(i)}=V.
	\end{equation}

	\sangria In this case, let’s recall that the particles do not interact electromagnetically with each other, rather they only interact with an electromagnetic field that permeates the whole Spacetime. Then we obtain the following Non-Relativistic Schr\"odinger equation:

	\begin{equation}\label{6}
	i\hbar\frac{\partial \tilde{\Psi}}{\partial T}=-\hbar^2\sum_{i=1}^{N}\frac{\nabla^{(i)2}\tilde{\Psi}}{2m_i}+V\tilde{\Psi}
	\end{equation}

	by adding the Equations \eqref{2} and using identities \eqref{3}, \eqref{4} and \eqref{5}.

	\sangria Equation \eqref{6} is the usual Non-Relativistic Schr\"odinger Equation, although it does not include an Electromagnetic Interaction between the particles in the potential $V$, only the interaction of the Particles with a Single External Electric Potential. 
	
	\sangria The Original Symmetry of the Beginning Multitemporal Wave Equation, where all the independent space and time variables of the particles were generic has been broken and in its stead we have only one Configurational Time $T$ and the generic space variables of all the particles. 
	
	\sangria Recapitulating, we took the Non-Relativistic Limit and at this Limit we managed to decouple the Configurational Time Variables from the space ones. We obtained a Conditional Probability Wave Function with Times equal to $\sigma$ plus a constant that is dependant on the considered index. We use some Arrangement Function to construct a $N$-space-time density whose Support is Spacelike. In this Support the Generic Coordinates of any two different particles have a Spacelike separation between them. This produces a single Schr\"odinger equation at the end, with a single Configurational Time. Then this last Wave Function will determine, taking its squared modulus, the Probability Density of finding the System of particles, according to the usual Born rule, at a single given configurational time $T$.
		
	\sangria The fact that the time in the Non-Relativistic Schr\"odinger equation is Configurational while the space variables are generic could explain why the uncertainty relation between Time and energy is different, in the Non-Relativistic Bohmian Mechanics, from the uncertainty relation between position and momentum. Maybe we could also explain why there is not operator associated to Time in the Non-Relativistic Limit and why it does not make sense to think of $i\hbar\frac{\partial}{\partial T}$ as an energy operator.

        \subsection{SYSTEM OF PARTICLES TYPE DIRAC} 
  
        \sangria The Dirac Equation for a particle is
        
        \begin{equation}\label{Dirac}
		 {\gamma_{\mu}} \left( p_{\mu} - \frac{e}{c} A_{\mu} \right) \Psi = -m c \Psi
		\end{equation}
		
		where
		
		\begin{equation}
		\gamma_i = 
		\begin{pmatrix}
			{\bf 0} & {\sigma_i} \\ {\sigma_i} & {\bf 0} 
    	\end{pmatrix}
    	\quad i=1,2,3
		\end{equation}
		
		and
		
		\begin{equation}
		\sigma_1 = 
		\begin{pmatrix}
		0 & 1 \\ 1 & 0 
		\end{pmatrix}
		\quad\sigma_2 = 
		\begin{pmatrix}
		0 & -i \\ i & 0 
		\end{pmatrix}
		\quad\sigma_3 = 
		\begin{pmatrix}
		1 & 0 \\ 0 & -1 
		\end{pmatrix}
		\end{equation}
		
		are the Pauli matrices. Besides
		
		\begin{equation}
		\gamma_4 = i \beta\quad\text{donde} \quad \beta =		
		\begin{pmatrix}
		1 & 0 & 0 & 0 \\ 0 & 1 & 0 & 0 \\ 0 & 0 & -1 & 0 \\ 0 & 0 & 0 & -1
		\end{pmatrix}
		\end{equation}
		
		\sangria If the spinor $\Psi$ is written by components we obtain $\psi_i$ and the equation \eqref{Dirac} becomes
			
		\begin{equation}
		({\gamma_{\mu}})_{ij} \left( p_{\mu} - \frac{e}{c} A_{\mu} \right)  {\psi}_{j} = - m c  {\psi}_{i} \quad \forall i 
		\end{equation}
		
		In the case of two particles we have $\psi_{ij} = \psi_{ij} \left( x^{(1)}_\mu, x^{(2)}_\nu \right) $
		
		\begin{equation}
		 ({\gamma_{\mu}})_{ij} \left( p_{\mu} - \frac{e}{c} A_{\mu} \right)  {\psi}_{jl} = - m c  {\psi}_{il} \quad \forall i,l 
		\end{equation}
		
		and
		
		\begin{equation}
		 ({\gamma_{\mu}})_{ij} \left( p_{\mu} - \frac{e}{c} A_{\mu} \right)  {\psi}_{lj} = - m c  {\psi}_{li} \quad \forall i,l
		\end{equation} 
        
        \sangria It can be proven that the expression $R^2 (x^{(1)}_\mu, x^{(2)}_\nu) = |\psi^*_{il}\beta _{lm}\beta _{ij}\psi _{jm}|$ (On the right hand side we use Einstein's notation of adding over repeated indices and where the symbol * denotes the complex conjugate) is transformed by passing from one frame of reference to another like a Scalar. In \cite{3} it was proposed to use this quantity as the space-time density for the two-particle system. It was found that using this choice a good Classical Limit is obtained. The following continuity equations were also found for each particle.
              
        $$ \frac{\partial \left( c\psi^*_{il}\beta_{lm}(\beta\gamma_\mu)_{ij}\psi_{jm}\right)}{\partial x^{(1)}_\mu} = 0 $$
    
        $$ \frac{\partial \left( c\psi^*_{il}\beta_{ij}(\beta\gamma_\mu)_{lm}\psi_{jm}\right)}{\partial x^{(2)}_\mu} = 0 $$.
    
    Following a procedure entirely analogous to the previous subsection, we obtain
    
    \begin{equation}\label{continuo}
    \frac{\partial R^2}{\partial\sigma} + \frac{\partial }{\partial x^{(1)}_\mu} \left(  R^2 V^{(1)}_\mu \right) +\frac{\partial }{\partial x^{(2)}_\mu}\left(  R^2 V^{(2)}_\mu \right) = 0
    \end{equation}
    
    where
    
    $$V^{(1)}_\mu=\frac{c\psi^*_{il}\beta_{lm}(\beta\gamma_\mu)_{ij}\psi_{jm}}{R^2}$$
    
    and
    
    $$V^{(2)}_\mu=\frac{c\psi^*_{il}\beta_{ij}(\beta\gamma_\mu)_{lm}\psi_{jm}}{R^2}.$$
    
    \sangria The equation \eqref{continuo} is the global continuity equation for this system.
    
	\sangria Let us consider, as in [\cite{3}], a Bohmian system, based on the Dirac equation, of two entangled particles. The guiding equations are
	
	$$\frac{dX^{(1)}_\mu}{d\sigma}=V^{(1)}_\mu$$
	
	$$\frac{dX^{(2)}_\mu}{d\sigma}=V^{(2)}_\mu$$

    where the right hand sides of these equations have been evaluated in the configurational coordinates of the particles.  At the Non-Relativistic Limit, it is usually considered that only the electronic components of the spinor are important. It is also needed to make a correction in the phases of the spinor.
    
    \sangria In the Non-Relativistic limit, we have
    
    \begin{equation}
    \begin{pmatrix}
    \psi_1 \\ \psi_2 \\ \psi_3 \\ \psi_4
    \end{pmatrix}
    \exp \left( {i \frac{c^2}{\hbar} m t} \right) \rightarrow
    \begin{pmatrix}
    \varphi_1 \\ \varphi_2 \\ 0 \\ 0
    \end{pmatrix}.
    \end{equation}
    
    \sangria In the case of two particles, we have the following Non-Relativistic Limit
    
    \begin{equation} 
    \begin{pmatrix}
    \psi_{11} & \psi_{12} & \psi_{13} & \psi_{14} \\ \psi_{21} & \psi_{22} & \psi_{23} & \psi_{24} \\ \psi_{31} & \psi_{32} & \psi_{33} & \psi_{34} \\ \psi_{41} & \psi_{42} & \psi_{43} & \psi_{44}
    \end{pmatrix}
    \exp \left(  i \frac{c^2}{\hbar} m \left( t^{(1)} + t^{(2)} \right) \right) \rightarrow
    \begin{pmatrix}
    \varphi_{11} & \varphi_{12} & 0 & 0 \\ \varphi_{21} & \varphi_{22} & 0 & 0 \\ 0 & 0 & 0 & 0 \\  0 & 0 & 0 & 0
    \end{pmatrix}.
    \end{equation}
    
     \sangria Doing so, the Guiding Bohmian Equations can be obtained. Now we occupy ourselves with studying the Bohmian Guiding Equations corresponding to the time coordinates: 

	$$ic\frac{dT^{(1)}}{d\sigma}=i\frac{c\varphi^*_{il}\delta_{lm}\delta_{ij}\varphi_{jm}}{|\varphi^*_{il}\delta_{lm}\delta_{ij}\varphi_{jm}|}$$
	
	$$ic\frac{dT^{(2)}}{d\sigma}=i\frac{c\varphi^*_{il}\delta_{ij}\delta_{lm}\varphi_{jm}}{|\varphi^*_{il}\delta_{lm}\delta_{ij}\varphi_{jm}|}$$
	
	\sangria When we take the Non-Relativistic Limit, the term inside the modulus in the denominator is positive and so we can remove the bars of absolute value. And we obtain
	
	$$\frac{dT^{(1)}}{d\sigma}=1$$
	
	$$\frac{dT^{(2)}}{d\sigma}=1$$
	
	and therefore
		
	$$T^{(1)}=\sigma+\delta^{(1)}$$
	
	$$T^{(2)}=\sigma+\delta^{(2)}$$.	
	
	\sangria We have two configurational times $T^{(1)}$ and $T^{(2)}$ . Let us now think about the Arrangement Function when the interaction is spacelike, exactly as the case of Klein-Gordon Equation. If the $2$-spacetime coordinates of the particles are in the Support of the Arrangement Function, then their interval is Spacelike. When taking the Non-Relativistic Limit we obtain, althought not everything has been represented here: two Pauli equations, one for each particle acting on the index of the Spinor and the Generic Coordinates Corresponding to the particle associated to this equation. Now, since the configurational times are $\sigma$ plus a constant and we must obtain that the interval between the particles is Spacelike, we have that
	
	\begin{equation}\label{desigualdad}
	|\delta^{(2)}-\delta^{(1)}|^2  \leq  ||X^{(2)}-X^{(1)}||^2/c^2.
	\end{equation}
	
	\sangria If we now make $c$ tends to infinity, the right side of the inequality tends to $0$ and so we obtain $\delta^{(2)}=\delta^{(1)}$ which means that $T^{(1)}=T^{(2)}=T$. The exact same procedure as before applies here to obtain a Single Wave Equation for both particles, which is a Spinor of two particles with spin $\frac{1}{2}$ with two indices and with the space coordinates of both particles. There the Guiding Bohmian Equations of the two particles describe how the space coordinates depend on the single configurational time coordinate. The Relativistic Multitemporal formalism yields in this case, at the Non-Relativistic Limit, the typical equation for two particles with spin $\frac{1}{2}$ with generic space coordinates and a single configurational time. Exactly as in the case of spinless particles.
	
	\begin{equation}
	 i \hbar \frac{\partial\varphi_{ij} }{\partial T} = \frac{1}{2m} \left( {\bf p}^{(1)} \cdot {\bf p}^{(1)} + {\bf p}^{(2)} \cdot {\bf p}^{(2)}  \right) \varphi_{ij} + (e {\phi}^{(1)} + e {\phi}^{(1)} ) \varphi_{ij} 
	\end{equation}      
	
	\sangria It is interesting too to study the SemiRelativistic Limit. In this case in the Inequality \eqref{desigualdad} We supose that $||X^{(2)}-X^{(1)}||^2/c^2 \leq \Delta^2 $ where $\Delta$ is lower than the resolution of our clocks and We obtain $\delta^{(2)}=\delta^{(1)}$, that is to say $T^{(1)}=T^{(2)} = T$. The Wave Equation is then

	$$i \hbar \frac{\partial\varphi_{ij} }{\partial T} = \frac{1}{2m} \left( \left( {\bf p}^{(1)} - {\frac{e}{c}} {\bf A}^{(1)} \right)^2 + 
	\left( {\bf p}^{(2)} - {\frac{e}{c}} {\bf A}^{(2)} \right)^2  \right) \varphi_{ij} + (e {\phi}^{(1)} + e {\phi}^{(1)} ) \varphi_{ij}$$ 
	$$- \frac{\hbar e}{2 m c} \left( {\bf B} \cdot {\bf\sigma} \right)_{il} \varphi_{lj} 
	- \frac{\hbar e}{2 m c} \left( {\bf B} \cdot {\bf\sigma} \right)_{jl} \varphi_{il}$$.

	\subsection{ANOTHER KIND OF ARRANGEMENT FUNCTIONS}
	
	
	\sangria Another different Arrangement Function  could be constructed by means of the same Function $F(u)$ that The Function that we used before in order to have interations esencialy spacelike. Now, We construct the Function
	
	$$ {G_{(i,j)}}(u) = 1 - \frac{F(-u) + F(u-b)}{2} $$
	
	$$ \mbox{ donde }  u =  ||x^{(i)}-x^{(j)}||^2-c^2(t^{(i)}-t^{(j)})^2  > 0 $$
	
	
	\sangria The Arrangement Function $ {G_{(i,j)}}(u) $ has the property that it has a $1$ value in the domain $ [0,b] $ and practically is null, 
	albeit of extremely small value outside this domain. That is to say, in reality, it is positive definite everywhere. That is very important because in the procedure that will be implemented we will divide by this function and this division has to be defined everywhere. 
	
	\begin{figure}[h!]
		\centering
	    \includegraphics[scale=0.5]{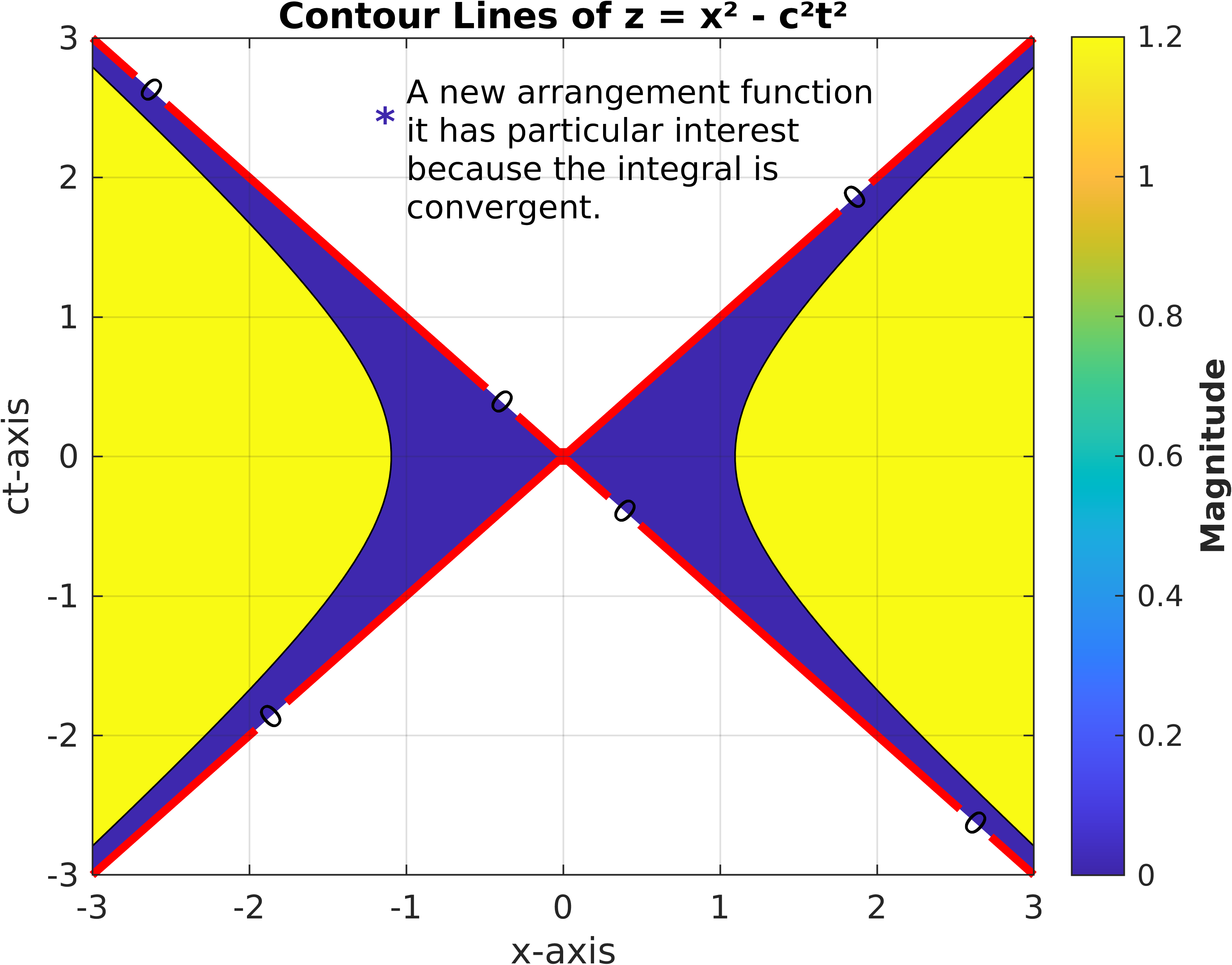}
	    \label{fig:arrngement}
	\end{figure}
	
	\begin{figure}[t!]
	     \centering
	     \includegraphics[scale=0.5]{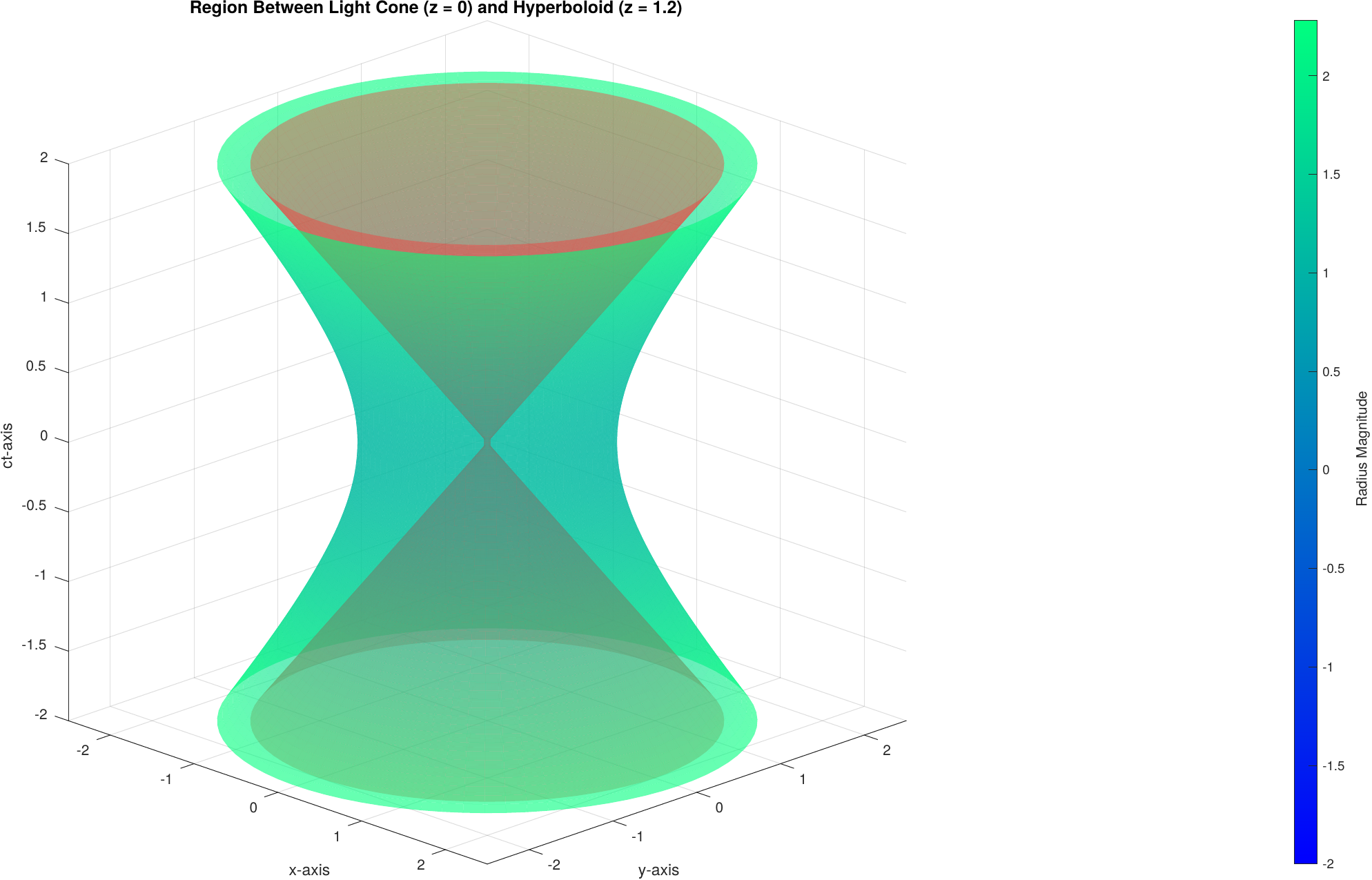}
	 \end{figure}
	
	\sangria We take now the Equations of Continuity (we call by the instant like that because we have not demonstrated that a Probability Density Function (PDF) exists for the Klein-Gordon Equations and the Dirac Equation. These are
	
	
	\begin{equation}
	\frac{\partial}{\partial {{x_\mu}^{(i)}}}  \left( R^2  \left( \frac{\partial S}{\partial {{x_\mu}^{(i)}}} -\frac{e}{c}{A_\mu}^{(i)}  \right) \right) = 0.
	\end{equation}
	
	
	\sangria For a system of two particles governed by Dirac Type Equations, we obtain the two following Continuity Equations:

	$$ \frac{\partial}{\partial x^{(1)}_\mu} { \left( c\psi^*_{il}\beta_{lm}(\beta\gamma_\mu)_{ij}\psi_{jm}\right)} = 0 $$

	$$ \frac{\partial}{\partial x^{(2)}_\mu} {\left( c\psi^*_{il}\beta_{ij}(\beta\gamma_\mu)_{lm}\psi_{jm} \right)} = 0 $$

	\sangria If we remember, $R^2 (x^{(1)}_\mu, x^{(2)}_\nu) = |\psi^*_{il}\beta _{lm}\beta _{ij}\psi _{jm}|$ is the space-time density proposed in the earlier Bohmian Models. In principle, one can try to rewrite the Equations in a different form.
	
	
	Klein-Gordon Equations:
	
	\begin{equation}\label{kleingordoncontinuidad}
	\frac{\partial}{\partial {{x_\mu}^{(i)}}}  \left( R^2 \frac{1}{m_i} \left( \frac{\partial S}{\partial {{x_\mu}^{(i)}}} -\frac{e}{c}{A_\mu}^{(i)}  \right) \right) = \frac{\partial}{\partial {{x_\mu}^{(i)}}}  \left( R^2  {V_{\mu}^{(i)}}  \right)  = 0.
	\end{equation}
	
	Dirac Equations:
	
	$$ \frac{\partial}{\partial x^{(1)}_\mu} { \left( R^2 \frac{{c\psi^*_{il}\beta_{lm}(\beta\gamma_\mu)_{ij}\psi_{jm}}}{R^2} \right)} = \frac{\partial}{\partial {{x_\mu}^{(1)}}}  \left( R^2  {V_{\mu}^{(1)}}  \right) = 0 $$

	$$ \frac{\partial}{\partial x^{(2)}_\mu} { \left( R^2  \frac{c\psi^*_{il}\beta_{ij}(\beta\gamma_\mu)_{lm}\psi_{jm}}{ R^2 } \right)} = \frac{\partial}{\partial {{x_\mu}^{(2)}}}  \left( R^2  {V_{\mu}^{(2)}}  \right) =  0 $$ 
	
	
	\sangria All the Continuity Equations can be written in this manner. Therefore, we study in general this kind of Continuity Equation. We consider, by the instant, two entanglement particles by a Relativistic Wave Function:
	
	$$ \frac{\partial}{\partial {{x_\mu}^{(1)}}}  \left( R^2  {V_{\mu}^{(1)}}  \right) = 0 $$

	$$ \frac{\partial}{\partial {{x_\mu}^{(2)}}}  \left( R^2  {V_{\mu}^{(2)}}  \right) =  0 $$ 
	
	\sangria $ G_{(1,2)}({x_\mu}^{(1)}, {x_\mu}^{(2)}) $ would be our Arrangement Function. Precisely, we arrange the two previous Continuity Equations  in the de following form:
	
	$$ \frac{\partial}{\partial {{x_\mu}^{(1)}}}  \left( G_{(1,2)} R^2 G_{(1,2)}^{-1}  {V_{\mu}^{(1)}}  \right) = 0 $$

	$$ \frac{\partial}{\partial {{x_\mu}^{(2)}}}  \left( G_{(1,2)} R^2 G_{(1,2)}^{-1} {V_{\mu}^{(2)}}  \right) =  0 $$ 
	
	\sangria We are going to consider that $ G_{(1,2)} R^2 $ is the 2-space-time density of the Models arranged in this form. By  the other side
	
	$$ \frac{d{X_{\mu}^{(1)}}}{d \sigma} = G_{(1,2)}^{-1}  {V_{\mu}^{(1)}} $$
	
	$$ \frac{d{X_{\mu}^{(2)}}}{d \sigma} = G_{(1,2)}^{-1}  {V_{\mu}^{(2)}} $$
	
	are the Bohmian Guiding Equations arranged with the procedure. In the Region where the Arrangement Function has the value $1$, we obtain the old Guide Equations published in the 2010 papers (Hern\'andez-Zapata and Hern\'andez-Zapata). Then, it is possible to take the two Limits, Classical and Non-Relativistic, that were worked in these papers without any problem. The question is if
	
	
	$$ G_{(1,2)} R^2 $$
	
	can be considered a Probability Density Function when it is normalized. The principal thing now, it must be integrable with finite integral. That is to say, the integral
	
	
	$$ \int_{space^{(1)}-time^{(1)}} \int_{space^{(2)}-time^{(2)}} G_{(1,2)} R^2  dV_{(1)} c dt_{(1)} dV_{(2)} c dt_{(2)}  $$
	
	must be finite. We considere that $ {x_\mu}^{(1)} $ is fixed and we make the integral over $ {x_\mu}^{(2)} $. That is to say, let's analize the partial integral

	
	$$ \int_{space^{(2)}-time^{(2)}} G_{(1,2)} R^2  dV_{(2)} c dt_{(2)} $$
	
	it would seem that this partial integral ought to have a finite value for $ {x_\mu}^{(1)} $ fixed. That is due to the manner in which the branches of the hyperbola $ ||x^{(i)}-x^{(j)}||^2-c^2(t^{(i)}-t^{(j)})^2 = 1.2 $ stick together to the Light Cone.
	
	
	
	 
	\sangria Since the integration Region becomes more compressed as we move away from the vertex of the Light-Cone, and the spacetime density tends to zero at the Light-Cone, this integral must converge over the Region between the Light-Cone and the fixed-value level curve of the integral (in Example 1.2). That is, it must converge over the unbounded branches of the 'Support of the Arrangement Function'. Like it is convergent in the bounded part of the Support of the Arrangement Function we conclude that the integral is convergent in all the 'Support of the Arrangement Function'. Like we are integrating over the space-time coordinates $ {x_\mu}^{(2)} $ and we obtain a finite value, we could think that the integral is already convergent. But we have forgotten  that what we have obtained is a function of the coordinates $ {x_\mu}^{(1)} $ and that it is necessary to integrate over these coordinates in all the space-time. There is not any warranty  that the integral in all the Configuration space-time of the two particles converge. That is to say, we can not normalize the Configuration space-time density of the two particles in order to convert it in a Probability  Density Function (PDF).
	
	
	\sangria 1.-To introduce an Arrangement Function is like to say that the Wave Function only has some sense in the domain where the Arrangement Function is relevant, that is to say, where the Arrangement Function equals 1. But the same argument can be set against the ideas of a Foliation, that is to say, a Spacelike hypersurface of 3-dimension that is the leaf of a foliation. Finally, by example, when 'The Bell School (Goldstein-Dürr-Zanghi-Tumulka)' establish the Bohmian velocities over the Foliation they only takes into account the values of the Wave Function over the Foliation and then It makes the same sins that us. Besides it uses properties of the leaf of the Foliation to evaluate the Bohmian Velocities over the leaf  of the Foliation. In our case, with the Arrangement Function, we utilize it to restrict the Domain where the $N$-space-time Density is Relevant. This helps us, all things considered, to obtain a $N$-space-time Density  that be integrable. We use the Arrangement Function to Evaluate the Bohmian Velocities, but in the zone where the Arrangement Function equals 1 these Bohmian Velocities do not be altered with respect to the Bohmian Velocities in the Not-Arrange Relativistic Bohmian Models. 
	
	
	\sangria 2.-The Arrangement Function introduced  by us does not cancel the Non-Locality in any way. By example, the hyperboloid that is around the Light Cone can have a very big diameter (1000 light-years or something like that). All the interactions happen  in this example and this Arrangement Function in a Spacelike manner. Now if we consider Cosmological Scales, the Non-Locality seems to be hidden and the Interactions in very big orders of magnitudes seem to happen  over the Light-Cone. In the ordinary physical scales that we use commonly, it is impossible to avoid the Consequences of the John Bell Theorem.
	
	
	\sangria 3.- The Arrangement that we did about the Born-Nikoli\'c rule produces that at least the partial integral over the space-time generic coordinates of a particle converge. Well, what we say is that the Arrangement Functions can serve to construct Bohmian Models Lorentz Covariant that can be utilized in practice and besides this practice becomes very much accesible to researchers that have competence in the Non-Relativistic Quantum Mechanics because the methods or Recherch become very similar. We have another task to accomplish, to show that efectively it is possible to obtain Lorentz Covariant Bohmian Models (Nikoli\'c, Hern\'andez-Zapata and Hern\'andez-Zapata) stationary respect to $ \sigma $ and where  it will be possible to construct a $N$-space-time probability density function and to explore Quantum extensions to Relativity.

	\subsection{A TENTATIVE SOLUTION}
	
	
	\sangria In the space-time we construct Timelike Lines. Over a Time Line, we take like a parameter the proper time of the Line. We define a Probability Density Function $ F( \tau ) $ (PDF) over the Line. And the Probability Density Function satisfies obviously
	
	$$ \int_{-\infty}^{\infty} F( \tau ) d\tau = 1  $$
	
	$ F( \tau ) $ is normalized. If we think in the Nikoli\'c Time $ \sigma $ then $ F( \tau ) d\tau $ is the probability that the particle that moves on the Timelike Line be found  in the proper-time interval $ d\tau $ over the Time Line. Each Time-Line $ \gamma_i $ provided with a Probability Density Function $ {F_i} ( \tau ) $ would be an element of the Complete Set of Time-Lines $ \Sigma $.
	
	
	
	Over each Time-Line a 1-dimensional dynamical exists, It is possible that a Wave Function over the Time-Line  exists and determines $ F( \tau ) $, at the end what we have is a Probability Density Function Unidimensional over the Time-Line.
	
	
	\sangria We think now in the Wave Function $ \Psi ( \vec{x}, t ) = R \exp({i {\frac{S}{\hbar}}}) $  for the Klein-Gordon Equation or the Spinor $\psi$ for the Dirac Equation. We remember that we can obtain a Continuity Equation for each type of Wave Equation.
	
	\sangria The two examples, (a) Klein-Gordon Equation and (b) Dirac Equation. In these two cases we construct the Continuity Equation
	
	(a)
	
	 \begin{equation}\label{continuidadKG}
	  \frac{\partial}{\partial {{x_\mu}}}  \left( \frac{R^2}{m_i}  \left( \frac{\partial S}{\partial {{x_\mu}}} -\frac{e}{c}{A_\mu}  \right) \right)=0
	\end{equation}
	
	(b)
	
	\begin{equation}\label{continuidadDirac}
	\frac{\partial}{\partial {x_\mu}} \left( c \bar{\psi} \gamma_\mu {\psi}  \right) = 0
	\end{equation}
	
	
	\sangria Each type of particle (a) and (b) get entangled to a Time-Line provided by a Probability Density Function $ F(\tau)$.
		
	\sangria We use for now the space-time Density $ R^2 $ for the Klein-Gordon Equation. 
	
	\sangria At the End All the Continuity Equations are
	
	$$ \frac{\partial}{\partial \tau} \left( F( \tau ) V_\tau \right) = 0 $$
	
	$$  \frac{\partial}{\partial {x_\mu}} \left( c \bar{\psi} \gamma_\mu {\psi}  \right) = 0 $$
	
	$$  \frac{\partial}{\partial {x_\mu}} \left( {\frac{R^2}{m}} \left( \frac{\partial S}{\partial {{x_\mu}}} -\frac{e}{c}{A_\mu}  \right) \right)=0 $$
	
	\sangria The first thing is that we are going to rewritten this Equations to apply the Bohm Ideas.
	
	$$ \frac{\partial}{\partial \tau} \left( F( \tau ) V_\tau \right) = 0 $$
	
	$$  \frac{\partial}{\partial {x_\mu}} \left( {\frac{R^2}{m}} \left( \frac{\partial S}{\partial {{x_\mu}}} -\frac{e}{c}{A_\mu}  \right) \right)=0 $$
	
	$$  \frac{\partial}{\partial {x_\mu}} \left( {\left| \bar\psi \psi \right| } \frac{c \bar{\psi} \gamma_\mu {\psi}}{ \left| \bar\psi \psi \right| }  \right) = 0 $$
	
	\sangria We will make some changes about the Continuity Equations and we will utilize the Arrangement Function whose 'Support' is situated between the Light Cone and the Contour Line 1.2 of the Interval Function. We call this Function $ G({x_\mu}, \tau) $. How can we visualize it? The Time-Line is parameterize by $ \tau $. Take a $\tau$ value and then we have an event on the Time-Line $ \gamma_i $. We then construct the Arrangement Function between Countour Lines 0 and 1.2 whose vertice is in the event of $ \gamma_i $ represented by $ \tau $. Then, over the Time-Line the particle interacts with the particle of Klein-Gordon Equation according to the Arrangement Function $ G $. It is like  the particle moves over the Time-Line 'would shepherd' to the Klein-Gordon particle. The Time-Like particle seems completely decoupled of the Klein-Gordon particle whereas there are both on the 'Support' of the Arrangement Function $G$. Only when the System is in the transition between $1$ and becoming negligible it must be reveled this strange entanglement. 
	
	
	$$  \frac{\partial}{\partial {x_\mu}} \left( F( \tau ) {\frac{R^2}{m}} \left( \frac{\partial S}{\partial {{x_\mu}}} -\frac{e}{c}{A_\mu}  \right) \right)=0 $$
	
	$$ \frac{\partial}{\partial \tau} \left( F( \tau ) R^2 V_\tau \right) = 0 $$
	
	
	\sangria We are going to utilize the Method of the Arrangement Function, this Function is equal in this case to     $ G( {x_\mu}, \tau ) $. We introduce then a number $ 1 $ in the parenthesis of the previous Continuity Equations. That is, we introduce on each parenthesis $ G G^{-1} $. From this point on, we will arrange all the system.
	
	$$  \frac{\partial}{\partial {x_\mu}} \left( G F( \tau ) {\frac{R^2}{m}} G^{-1} \left( \frac{\partial S}{\partial {{x_\mu}}} -\frac{e}{c}{A_\mu}  \right) \right)=0 $$
	
	$$ \frac{\partial}{\partial \tau} \left( G F( \tau ) R^2 G^{-1} V_\tau \right) = 0 $$
	
	
	\sangria Then we arrange the terms appears a Space-time-$ \tau $ Density that is stationary respect to $ \sigma $, that is to say, it does not depend on $ \sigma $.
	
	$$ {\rho_{\text space-time-\tau }} = G F( \tau ) R^2 $$
	
	\sangria We arrange too the Bohmian Guide Equations, one in the 1-Space-time and the other over the Time-Line $ \gamma_i $:
	
	$$ \frac{d {X_\mu}}{d \sigma} = G^{-1} \frac{1}{m} \left( \frac{\partial S}{\partial {{x_\mu}}} -\frac{e}{c}{A_\mu}  \right) $$
	
	$$ \frac{d\tau}{d\sigma} = G^{-1} V_\tau   $$
	
	
	\sangria The first of the Equations is identical to our old Bohmian Equation Guide provided that the Configuration system remains inside the 'Support' of the Arrangement Function. The second Equation is a new Bohmian Guide Equation in one dimension that describes the mouvement respect to $ \sigma $ of one particle moving over one Time-Like Line set in Advance. Our Intuition is that these Lines must be geodesics in the correspondent Geometry. In this case Plane because we are treating Special Relativity, then the $ \gamma $ lines must be geodesics, that is right trajectories in space-time. Besides, we do not suppose many properties in these particles. The only role that these particles would be following geodesics of the given Geometry of space-time and to shepherd to the Klein-Gordon particles. Therefore the only physical property would be the mass. We suppose that this is precisely the characteristic of Obscur Matter. Something that has mass and therefore gravity but it can not emit any radiation of other type. If this argument is good it would be interesting because it means that the Obscur Matter, instead of being an arbitrary added to the Nature, it would be a Fundamental Component in the Constitution and Fundamentation of the Laws of Nature itself.
	
	\sangria It is neccesary to study the Behavior of the space-time-$ \tau $ density to see if this time we can construct a Probability Density Function.
	
	
	
	\sangria In this Section we have seen  the case of a Klein-Gordon particle and a Time-Like Line particle that shepherds the first. The second particle moves in a geodesic in the space-time, Plane in Special-Relativity. The problem of a System of several entanglement particles of this type will be treated later.
	
	\subsection{Convergence of the Integral}
	
    $$ C_{\text{All Configuration}} = \int_{-\infty}^{\infty} \int_{All-space-time} {\rho_{\text space-time-\tau }} c dt dV d \tau $$
    
    $$ = \int_{-\infty}^{\infty} \int_{All-space-time} G F( \tau ) R^2 c dt dV d \tau $$
    
    $$ = \int_{-\infty}^{\infty} F( \tau ) \left( \int_{All-space-time} G R^2 d V c d t \right) d \tau $$
    
    \sangria We have the following estimation
    
    $$ \int_{All-space-time} G R^2 d V c d t = \int_{\text{Support of G for fixed } {\tau}} R^2 d V c d t = A (\tau) $$
    
    $$ C_{\text{All Configuration}} = \int_{-\infty}^{\infty} F( \tau ) A(\tau ) d \tau = \bar{A} $$
    
    \sangria We will explore if $ \bar{A} $ is finite. 
    
    \sangria In order to study this problem that has a very important relation with the convergence that we look for, we are going to point out a few key points to highlight.

    \sangria Regarding the Time-like trajectory of the particle (whose motion is unidimensional and on this trajectory its position is fixed by the proper time), we choose an interval, which can be any interval we want [a, b], and afterwards we build an 'Arrangement' Function that is negligible outside the interval [a, b], and furthermore that is of class $ C^{( \infty )} $. Then, the average value of any function on this interval is finite. First, let us consider the integral over $ x_{\mu} $ as this Arrangement Function is the gap between the level curves 0 and 1.2 of the Interval Function, assuming that the vertex is found on the trajectory identified by $ \tau $. Then the integral over $ x_{\mu} $ produces a certain value $ A(\tau) $, so we have a normalized function for a time density, and since $ F (\tau) $ is an Arrangement  Function with typical characteristics of the Arrangement Functions that we have been using, we must have that:
    
    $$ \int_{-\infty}^{\infty} A( \tau ) F(\tau ) d \tau =  \bar{A} $$
    
    which must be finite, and therefore, the spacetime-$ \tau $ Density of the System is integrable. If we integrate, we finally obtain the normalization condition, so in the end we have a Probability Density Function (PDF) as we want. It was necessary to increase by one the configuration space but we manage to extend the idea of a Born density of Non-Relativistic Quantum Mechanics to Relativistic Bohmian Mechanics Models type Nikoli\'c, Hern\'andez-Zapata.
    
    \sangria PHYSICAL OBSERVATION: It is fundamental to clarify the nature of the system outside the interval [a, b]. In these prior and subsequent regions, the particle lacks an existence in the strictly  Bohmian (or traditional quantum) sense. This arises because its dynamics are not governed by the standard guiding factor derive from the wave function, but rather by the inverse of the Arrangement Function introduced in the formalism. Consequently, the particle's entry into the interval [a, b] formally represents the emergence or "creation" of the systems's quantum  phenomenology, where the habitual Bohmian  guiding law becomes dominant. Similarly , its exit from said interval corresponds to the dissolution or "annihilation" of this quantum behavior, marking a transition toward a non-conventional regime where the standard Bohmian description is no longer applicable.

	\section{CONCLUSIONS}

	\sangria In this paper we have attempted to approach some ideas needed to give a fully fledged form to the Relativistic Covariant Bohmian Models type Nikoli\'c, Hern\'andez-Zapata. On the other hand, we have tried to polish the Non-Relativistic Limit in the two types of Models we have studied more carefully, one based on the Klein-Gordon Equation [\cite{1},\cite{2}] and the other based on the Dirac Equation [\cite{3}]. For this, \textit{“the Spacelike Interaction constructed by means of Arrangament Functions”} on the Wave Function was needed [\cite{5}]. Since this proposition greatly restricts the Subdomain of the Wave Function where it is Non-Zero, one could expect for the integral of the probability density function to be finite. 
	
	\sangria Working with the hypothesis of Space-like Interaction means that, while the Classical Interactions are strictly Time-like, the Quantum interactions are strictly Space-like. This kind of interactions are the responsible of Non-local behavior in Quantum Mechanics. We have shown in this article some consequences of the assuming the hypothesis about the Classical Limit in Relativistic Bohmian Mechanics Models. We have obtain a good Non-Relativistic Limit in which a singular configurational time is obtained and the Wave Function dependes on this Configurational time and on the generic spatial coordinates of all particles. The simetry between spatial coordinates and time coordinates of the Original Relativistic Model has been broken in the Non-Relativistic Limit. 
	
	\sangria Once we have made this kind of restriction using Arrangement Functions and the Procedure explained in the paper to arrange the Bohmian Model, We have explored the posibility to introduce for each Klein-Gordon (or Dirac) particle, an aditional particle with a Dynamics restricted to a geodesic in the Space-time Geometry (one dimensional manifold) and to study the entanglement between the particle in the four Dimensional space-time (Klein-Gordon or Dirac) and the particle moving in the one Dimensional Manifold. This was realized by means of Arrangement Functions. Like the new particle moving in the Geodesic must have only mass to remain in this one-dimensional manifold. We suppose that it is a particle of Obscur Matter. 
		
    \section{ACKNOWLEDGMENTS}
	
	\textit {We, the authors, would like to express our deepest gratitude to Ernesto Hern\'andez-Zapata, with whom this research line originally began. We also thank Patrice Le Gal, IRPHE (Marseille, France), for introducing us to the specific experiment of a ludion oscillating in a stratified fluid, proposing it as a manifestation of guiding dynamics. In particular, Sergio Hern\'andez-Zapata thanks Patrice Le Gal for keeping him motivated and working on experimental problems, such as macroscopic fluid analogies with Quantum Mechanics, which indirectly led him to reconsider the problematic nature of Bohmian Mechanics, a topic whose work he had put off for a long time. We consider that this paper fits well, even if only as an analogy, within the subject of fluids. The work on the present paper was supported by DGAPA-UNAM under project IN-113621 (Transporte de part\'iculas, convecci\'on y vorticidad). Un agradecimiento (entre mexicanos): Sergio Hern\'andez-Zapata quiere agradecer el humanismo de su profesora Catalina Stern Forgach (la cual me ense\~no mucho de lo que s\'e de Inestabilidades Hidrodin\'amicas), quien mand\'o llamar a una Ambulancia para que me llevaran a Atenci\'on de la UNAM debido a un colapso nervioso ocurrido en plena clase por 2009-2010. Y tambi\'en quiero agradecerle haberme dado permiso de separarme unas semanas de mis labores, lo que me permiti\'o avanzar mucho en la escritura de los art\'iculos bohmianos de 2010.}


\end{document}